%% file: main.tex
\documentclass[reprint,prr,twocolumn,showpacs,superscriptaddress,aps,longbibliography]{revtex4-2}

\usepackage{amsmath}
\usepackage{epstopdf}
\usepackage{amssymb}
\usepackage{amsfonts}
\usepackage{float}
\usepackage[colorlinks=true,citecolor=blue]{hyperref}
\usepackage{graphicx,graphics}
\usepackage{bm}
\usepackage{xcolor}
\usepackage{graphicx}
\usepackage{bm}
\usepackage{hyperref}
\usepackage{braket,bbold,color} 
\usepackage{tcolorbox}
\usepackage{ifthen}
\usepackage{dsfont}
\usepackage{tikz}
\usepackage{MnSymbol}
\usepackage{placeins}
\usepackage{quantikz}
\usepackage{graphicx}
\usepackage{subfigure}
\usepackage[percent]{overpic}
\usepackage{soul}

\newcommand{\modification}[1]{\textcolor{black}{#1}}

\begin{document}


\title{Quantum circuit algorithm for topological invariants of second order topological many-body quantum magnets}

\author{Sebastián Domínguez-Calderón}
\affiliation{Department of Applied Physics, Aalto University, 00076 Aalto, Finland}
\affiliation{School of Natural Sciences, Technische Universität München, D-85748 Garching, Germany}

\author{Marcel Niedermeier}
\affiliation{Department of Applied Physics, Aalto University, 00076 Aalto, Finland}

\author{Jose L. Lado}
\affiliation{Department of Applied Physics, Aalto University, 00076 Aalto, Finland}

\author{Pascal M. Vecsei}
\affiliation{Department of Applied Physics, Aalto University, 00076 Aalto, Finland}
\affiliation{Institute of Mechanical Engineering and Energy Technology, Lucerne University of Applied Science and Arts, CH-6002 Lucerne, Switzerland}

\date{\today}

\begin{abstract}
Topological quantum matter represents a flexible playground to engineer unconventional excitations. While non-interacting topological single-particle systems have been studied in detail, topology in quantum many-body systems remains an open problem. 
Specifically, in the quantum many-body limit, one of the challenges lies in the computational complexity of obtaining the many-body ground state and its many-body topological invariant. While algorithms to compute ground states with quantum computers
have been heavily investigated, 
algorithms to compute topological invariants in a quantum computer are still under active development. 
Here we demonstrate a quantum circuit to compute the many-body topological invariant of a second-order topological quantum magnet encoded in qubits. Our algorithm relies on a quantum circuit adiabatic evolution in transverse paths in parameter space, and we uncover hidden topological invariants depending on the traversed path. Our work puts forward an algorithm to leverage quantum computers to characterize many-body topological quantum matter.
\end{abstract}

\maketitle

\section{Introduction}

\input{intro}

\section{Methods}

\input{methods}

\section{Topological invariant of first order 1D quantum magnets}

\input{1d-case}

\section{Topological invariant of second order 2D quantum magnets}
\label{sec:2d-case}

\input{2d-case}

\section{Conclusion}

In summary, we have presented a quantum algorithm that
allows us to calculate topological invariants of a higher-order
topological quantum magnet. Our algorithm relies on a high-dimensional
path-dependent adiabatic evolution of the Berry phase,
enabling the calculation of topological
invariants in two-dimensional quantum many-body models.
We uncover a hidden behavior of the invariant for different paths through the HOSPT parameter space, suggesting that gauging the symmetry has intricacies and one has to be careful how to define the invariant, a result which had been previously overlooked. Our results make a first step towards leveraging quantum computers to rationalize
topology in many-body systems, including 
the nature of the quantum geometry formed by such parameter spaces of gauged HOSPT phases. 
It is finally worth noting that in the current formulation, 
the practical use of our algorithm requires substantial circuit depths
and accurate ground state preparation, which
are beyond the capabilities of currently available quantum computers.
Our results highlight the potential for using future fault-tolerant or topological quantum computers to calculate topological invariants of HOSPT phases.

\appendix
\section{Hadamard Test} \label{appendix:hadamard-test}
We focus on the use of the Hadamard test to extract the phase $\varphi$ produced by a unitary matrix $U$ on an eigenvector $\ket{\psi}$ \cite{Cleve1998}

\begin{equation}
    U \ket{\psi} = e^{i\varphi} \ket{\psi}.
\end{equation}
The Hadamard test consists of preparing an auxiliary qubit in a superposition state $\ket{\text{aux}} = \ket{+} = \frac{1}{\sqrt{2}} \left( \ket{0} + \ket{1} \right)$, then performing the controlled unitary $CU$ operation onto a test system prepared in the eigenstate $\ket{\psi}$, giving

\begin{equation}
    CU \ket{+} \otimes \ket{\psi} = \frac{1}{\sqrt{2}} \left( \ket{0} + e^{i\varphi}  \ket{1} \right) \otimes \ket{\psi}.
\end{equation}
The phase is now part of the auxiliary qubit's superposition, an effect named phase-kickback

\begin{equation}
    \ket{\text{aux}} \rightarrow \frac{1}{\sqrt{2}} \left( \ket{0} + e^{i\varphi}  \ket{1} \right).
\end{equation}
Then we return the auxiliary qubit to the computational basis via another Hadamard gate

\begin{equation}
    H \ket{\text{aux}} \rightarrow \frac{1}{2}\left(1 + e^{i\varphi} \right) \ket{0} + \frac{1}{2}\left(1 - e^{i\varphi} \right) \ket{0},
\end{equation}
and from the measurement statistics it is found $P(\sigma_z = +1) = \cos^2 (\varphi /2)$. This process is schematically shown in Fig. ~\ref{fig:hadamard-test}b to extract the Berry phase.

\section{Qiskit unitary gates}
\label{appendix:qiskit-gates}
The unitary operator for $H_0$ (cf Eq.~\ref{eq:unitary-h0}) can be implemented by Qiskit's $R_{XX}$, $R_{YY}$, $R_{ZZ}$ two qubit gates

\begin{align}
    &e^{-\frac{i}{\hbar} \frac{\Delta t}{2} H_0} = 
    \\
    &\prod_{i \, \in \, \text{odd}} 
    e^ {-\frac{i}{\hbar}  \frac{\Delta t}{2} J_1 \, \boldsymbol{S}_i \, \cdot \, \boldsymbol{S}_{i+1}} =
    \notag
    \\
    &\prod_{i \, \in \, \text{odd}} 
    e^ {-\frac{i}{\hbar}  \frac{\Delta t}{2} J_1 \, S^x_i S^x_{i+1}}
    e^ {-\frac{i}{\hbar}  \frac{\Delta t}{2} J_1 \, S^y_i S^y_{i+1}}
    e^ {-\frac{i}{\hbar}  \frac{\Delta t}{2} J_1 \, S^z_i S^z_{i+1}}
    \notag,
\end{align}
since $R_{XX}(\theta) = e^ {-i \frac{\theta}{2} \sigma_x \otimes \sigma_x}$, $R_{YY}(\theta) = e^ {-i \frac{\theta}{2} \sigma_y \otimes \sigma_y}$, $R_{ZZ}(\theta) = e^ {-i \frac{\theta}{2} \sigma_z \otimes \sigma_z}$. These gates will also work for $H_1''(t)$ (cf Eq.~\ref{eq:unitary-h1-2}). As for $H_1'(t)$ (cf Eq.~\ref{eq:unitary-h1-1}), Qiskit's $R_{XX+YY}$ two qubit gate is used

\begin{multline}
    R_{XX+YY}(\theta, \beta) = \\ 
    \left( e^{i \frac{\beta}{2} \sigma_z } \otimes I  \right)
    e^ {-i \frac{\theta}{4} \left( \sigma_x \otimes \sigma_x + \sigma_y \otimes \sigma_y \right)}
    \left(e^{i \frac{-\beta}{2} \sigma_z } \otimes I  \right).
\end{multline}
By setting $\beta = -\pi/2$, we can implement $H_1'(t)$, since

\begin{equation}
    R_{XX+YY}(\theta, -\pi/2) =  e^ {-i \frac{\theta}{4} \left( \sigma_x \otimes \sigma_y - \sigma_y \otimes \sigma_x \right)}.
\end{equation}
Note that the definitions follow from Qiskit v2.3.1, and might not hold for future versions.

\section{Decoupled energy degeneracy of 2D model}
\label{appendix:degen}
Starting from the 2D dimerized Heisenberg model described in Sec. ~\ref{sec:2d-case}, upon setting $J_2=0$, the system can be described as as a sum of plaquettes

\begin{equation}
    H= \sum_p H_p,
\end{equation}
where $H_p$ is just the Heisenberg spin-1/2 Hamiltonian for the plaquette $p$. The full energy spectrum will be a sum of the energy spectrums of all plaquettes. Thus, if one plaquette becomes degenerate, the whole spectrum becomes degenerate. By adding twists onto one plaquette, it's Hamiltonian will be

\begin{align}
    H_p = &\,J_1 \sum_{\braket{i,j}} \cos{\varphi_{ij}} \left( S^x_i S_j^x + S^y_i S_j^y \right) 
    \\ &- \sin{\varphi_{ij}} \left( S^x_i S_j^y - S^y_i S_j^x \right) + S^z_i S_j^z 
    \notag,
\end{align}
which only consists of four spins and can then be exactly diagonalized via analytical software such as Mathematica, to find that the lowest energy levels are given by Eq. ~\ref{eq:degen-energy}.

\textbf{Acknowledgments}
We acknowledge the computational resources provided by the Aalto Science-IT project. We acknowledge the support from the Research Council of Finland through grants (Grant No.~370912), 
the European Research Council through ERC-CoG grant ULTRATWISTROICS (No. 101170477),
InstituteQ, the Finnish Quantum Flagship, and the Jane and Aatos Erkko Foundation. We thank C. Flindt for useful discussions.

\bibliography{Ref-Lib}

\end{document}

%% file: intro.tex
With the advent of topological materials
 and their vast array of new physical phenomena \cite{Moore2010, wen2017colloquium}, 
 intense research has been devoted to their understanding and classification. Among them,
 symmetry protected topological (SPT) phases hold promise for engineering fault-tolerant quantum computing and spintronic devices \cite{Moore2010, Chiu2016, GonzlezCuadra2022}. SPT phases are hard to distinguish from trivial phases due to their subtle nature \cite{wen2017colloquium}, which means that they are generally not detectable by local order parameters and have no obvious distinction merely based on symmetry arguments \cite{Chiu2016}. SPTs are found in free-particle systems as well as in strongly-interacting matter such as topological crystalline quantum magnets \cite{Watanabe2017}. 1D and 2D gapped SPT phases have a full classification using group cohomology theory, and higher-dimensional generalizations have been developed \cite{chen2011twodimensional, Chen2011gapped1d}. While SPT phases in non-interacting systems are well understood \cite{chen2011twodimensional, Chen2011gapped1d, Wen2012}, many-body interacting systems remain a substantial open challenge \cite{Chiu2016, Wang2020}. 

The discovery of higher-order topological phases in non-interacting fermions enabled new exotic phenomena such as corner and hinge states~\cite{Benalcazar2017, PhysRevB.96.245115, Song2017, Langbehn2017}, which generalize the bulk-boundary correspondence of conventional topological insulators. Higher-order topological phases now have exhaustive classification for non-interacting systems~\cite{Schindler2018}. Theoretical models in interacting systems have also been shown to exhibit phenomena similar to higher-order topological phases \cite{dubinkin2019higherorder}. 
Higher-order symmetry-protected topological (HOSPT) phases can be built from lower-dimensional SPTs in a systematic way, and to a large extent can be classified using group cohomology \cite{Rasmussen2020}, which provides frameworks for calculating topological invariants using algebraic topology and cocyles of group cohomology \cite{Hung2014, Wen2017}. 
HOSPT states appearing in interacting bosonic and fermionic systems can exhibit corner modes~\cite{you2018higherorder},
and have attracted substantial attention over the last few years~\cite{song2017topological, Tamiya2021, kariyado2018Znberry, Dubinkin2023, Deng2024,  MayMann2022, You2020, Wienand2022, You2018, peng2021deconfined, Azses2023}.

Despite the existence of theoretical and computational methods to calculate topological invariants in non-interacting systems, the many-body case is an exponentially hard problem~\cite{Ayral2023}. Quantum computers are a natural candidate in the calculation of topological invariants of many-body Hamiltonians,
for which a variety of algorithms to compute ground states have been developed \cite{Liu2025, Yoshioka2025, Wille2017}.
When dealing with a topological system, not only the ground state, but also its topological
invariant becomes a crucial object to be computed \cite{Hung2014, Ahn2019}. This has motivated a variety of quantum circuit algorithms
for topological phases of matter in both non-interacting and interacting systems~\cite{Niedermeier2024, Murta2020, Kudo2019, Smith2022, Zhang2022, Koh2024, Sun2023}. As such, quantum computers hold potential for classifying, computing and predicting many-body topological quantum matter, and, specifically, interacting HOSPT phases.

In this work, we present a quantum circuit to compute topological invariants in higher-order symmetry protected topological phases. Specifically, we demonstrate our algorithm for computing the quantized Berry phase of a first-order, as well as a second-order SPT quantum magnet, both of which are described by a Heisenberg spin-1/2 model. Further models that could be studied with the same quantum circuit include the XX and XY models \cite{dubinkin2019higherorder}. The 2D tetramerized case exhibits a plaquette structure of bonds (Fig.~\ref{fig:tetra-model}), while the 1D dimerized case exhibits an alternating link structure (Fig.~\ref{fig:dimerized-results}). For such systems with anti-unitary symmetry (in this case time-reversal), the Berry phase topological invariant is quantized to $0$ or $\pi$, corresponding to a trivial phase and a HOSPT phase, respectively. We calculate topological invariants by simulating time-evolution via gate-based adiabatic quantum circuits, and extracting the Berry phase by means of a Hadamard test. Our results show the use case of quantum computing to classify higher-order symmetry protected topological phases, highlighting their potential for finding new phases of matter.

%% file: methods.tex
\begin{figure}[t]
    \centering
    \subfigure{
        \begin{overpic}[width=0.45\linewidth]{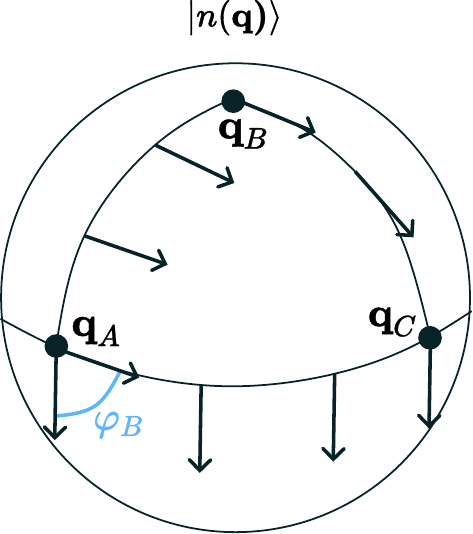}
            \put(-54,100){\small (a)}   
        \end{overpic}
    }
    \subfigure{
        \begin{overpic}[width=0.97\linewidth]{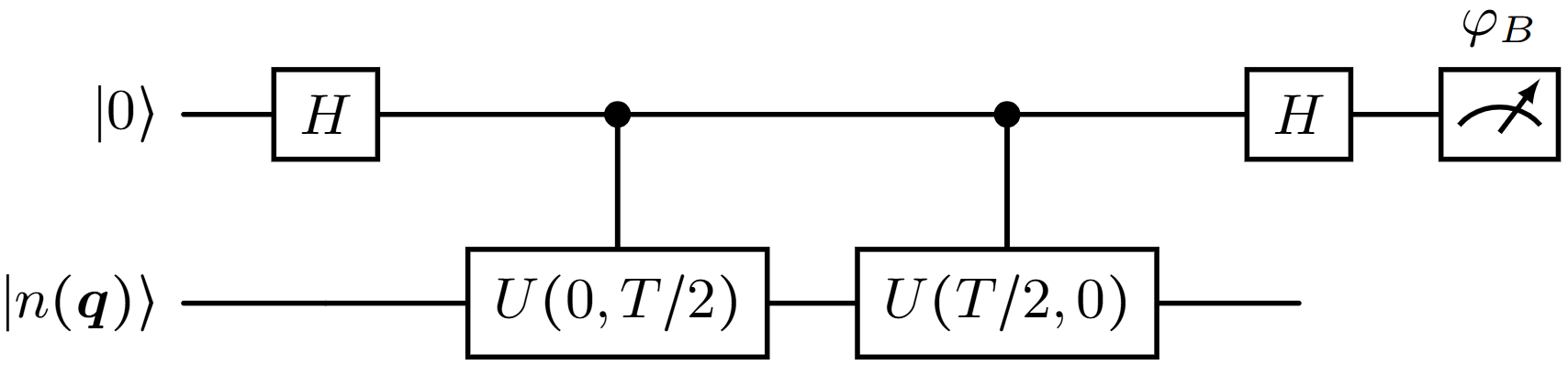}
            \put(-2,25){\small (b)}   
        \end{overpic}
    }
            
    \caption{(a) An instantaneous eigenstate evolved along a closed parameter path will pick up a Berry phase $\varphi_B$, which is a measure of how the complex vector changes along the path. This effect is analogous to parallel transport of vectors in curved geometry. (b) Proposed quantum circuit with a Hadamard test on an adiabatic time evolution $U(t_0, t_1)$ which allows to measure $\varphi_B$.}
    \label{fig:hadamard-test}
\end{figure}

To detect topological phase transitions, we exploit specific geometric phases
emerging during a cyclic evolution in a given parameter space. 
The Berry phase emerges as a general solution of adiabatic state evolution of a parametrized Hamiltonian $H(\boldsymbol{q})$, dependent on a set of parameters $\boldsymbol{q} = \left(q_1, q_2, \dots,q_n \right)$ \cite{Solem1993, Xiao2010}, which without loss of generality are defined within a circle, $q_i \in [0,2\pi]$. For each set of parameters, there exists a complete set of eigenstates such that

\begin{equation}
    H(\boldsymbol{q})\ket{n(\boldsymbol{q})} = E_n(\boldsymbol{q}) \ket{n(\boldsymbol{q})}.
\end{equation}
According to the adiabatic theorem, if the system is initialized in an eigenstate $\ket{n(\boldsymbol{q})}$, then a slow evolution of the Hamiltonian parameters will keep the state vector in the instantaneous eigenstate $\ket{n(\boldsymbol{q}(t))}$ at all times. Considering an adiabatic time evolution where parameters follow a closed path, eigenstates will pick up a total phase consisting of the difference between the dynamical phase $\varphi_D$ and the geometrical (Berry) phase $\varphi_B$

\begin{equation}
    U(0, T) \ket{n(\boldsymbol{q})} = e^{-i (\varphi_D - \varphi_B)}\ket{n(\boldsymbol{q})},
\end{equation}
where $U(0, T)$ denotes a time evolution from time $0$ to $T$,

\begin{equation}
    U(0, T) = \mathcal{T} \left\{ e^{-\frac{i}{\hbar}\int_0^T dt \, H(\boldsymbol{q}(t))} \right\},
\end{equation}
where $\mathcal T$ is the time ordering operator and the phases $\varphi_D$, $\varphi_B$ are given by

\begin{equation}
    \varphi_D = \int_0^Tdt \, E_n(\boldsymbol{q}(t)),
\end{equation}

\begin{equation}
    \varphi_B = i \int_0^T dt \, \bra{n(\boldsymbol{q}(t))} \partial_t \ket{n(\boldsymbol{q}(t))} .
\end{equation}
The closed loop adiabatic evolution of an eigenstate, given by a parametrized Hamiltonian, can be visualized as a statevector adiabatically moving along a curved space (Fig ~\ref{fig:hadamard-test}a). When returning to the original parameters, the vector may pick up a geometric phase $\varphi_B$.

Considering a general anti-unitary operator $\Phi = K U$, where $K$ denotes complex conjugation and $U$ is some unitary operator, it holds that its action on the Berry phase is $\Phi:\varphi_B \rightarrow -\varphi_B$. If the Hamiltonian in consideration has anti-unitary symmetry $[H, \Phi] = 0$, then the Berry phase must respect the symmetry $\varphi_B = -\varphi_B$. However, it is known that the Berry phase must be gauge invariant within a circle (modulo $2 \pi$). Both gauge invariance and anti-unitary symmetry quantize the Berry phase to $\varphi_B = \left\{ 0, \pi \right\}$ \cite{Hatsugai2006, Hatsugai2007, Hatsugai2011}.

To isolate the Berry phase, we can evolve the system back in time to cancel the dynamical phase, since it is anti-symmetric in time $\int_T^0 dt \, E(\boldsymbol{q}(t)) = - \varphi_D$ \cite{Murta2020}. The total phase picked up by an instantaneous eigenstate, after evolving forwards and then backwards in time, is twice the Berry phase

\begin{equation}
    U(T, 0)  U(0, T)  \ket{n(\boldsymbol{q})} = e^{i 2\varphi_B} \ket{n(\boldsymbol{q})},
\end{equation}
which is understood as a consequence of double looping the parameter space and the dynamical phase being canceled. For time-reversal symmetric Hamiltonians (also an anti-unitary symmetry), time evolution can be cut into half periods, such that time goes forward and backwards while the parameters perform only a single loop, allowing to distinguish the quantization $\varphi_B = \{ 0, \pi \}$ \cite{Murta2020} as

\begin{equation}
    \label{eq:half-time}
    U(0, T/2) U(T/2, 0)  \ket{n(\boldsymbol{q})} = e^{i\varphi_B} \ket{n(\boldsymbol{q})}.
\end{equation}

We implement such an adiabatic time evolution in a quantum circuit by dividing the Hamiltonian into $k$ parts such that $H(t) = \sum^k_{j=1} H_j(t)$, where each sub-Hamiltonian $H_j$ consists of only commuting operators. We then cut up time evolution into a sequence of small steps following the second-order Trotter-Suzuki decomposition \cite{Hatano2005, SUZUKI1993}

\begin{multline}
    \label{eq:trotter-suzuki}
    U_2(t + \Delta t) 
    \approx 
    e^{-\frac{i}{\hbar} \frac{\Delta t}{2}H_1 \left( t + \frac{\Delta t}{2} \right) }
    \dots 
    e^{-\frac{i}{\hbar}\frac{\Delta t}{2}H_{k-1} \left( t + \frac{\Delta t}{2} \right)}
    \\
    \times 
    e^{-\frac{i}{\hbar} \Delta t H_{k} \left( t + \frac{\Delta t}{2} \right)}
    e^{-\frac{i}{\hbar}\frac{\Delta t}{2}H_{k-1} \left( t + \frac{\Delta t}{2} \right)}
    \dots
    e^{-\frac{i}{\hbar} \frac{\Delta t}{2}H_1 \left( t + \frac{\Delta t}{2} \right) },
\end{multline}
such that evolution is then given by a product of small time steps

\begin{equation}
    U(0, T) \approx \prod_{j=0}^{N-1} U_2(t_j + \Delta t) 
\end{equation}
with $N$ the number of time steps, $\Delta t = T / N$, and $t_j = j \Delta t$. To extract the Berry phase from the circuit, we simply perform a Hadamard test on an auxiliary qubit (Fig.~\ref{fig:hadamard-test}b) \cite{Niedermeier2024, Cleve1998}, where the Berry phase is extracted from the measurement statistics of the auxiliary qubit in the computational basis as

\begin{equation}
    \label{eq:pop-statistics}
    P(\sigma_z=+1)=\cos^2 \left( \varphi_B/2 \right).
\end{equation}
An outline regarding the procedure of the Hadamard test can be found in App. ~\ref{appendix:hadamard-test}. 

\modification{
We simulate this quantum circuit by means of Qiskit's AerSimulator ~\cite{qiskit, aer}.
The typical usage, which we leverage, is to compile and simulate the quantum circuit on an initial state vector, and perform measurements on it. We use the ideal simulator without presence of noise. Our usage 
thus does not focus on a specific architecture of real (NISQ) devices, and disregards noise channels. However, the AerSimulator does have the capacity to compile the circuits in a way that the resulting sequence of gates respects existing architectures, for example to run it on IBM's chips. It would be further possible to consider the effects of various noise channels in the simulation, to see how robust the algorithm would be in the presence of decoherence mechanisms. So in this sense, the simulator can respect effects of real NISQ devices. This, however, is not the focus of our work, as we present and test a quantum algorithm to evaluate topological invariants of non-trivial many-body quantum magnets. 
As we elaborate in the following sections, the  circuits are too deep to reliably run on
NISQ devices with the currently available fidelities.}

In the following sections, the outlined methods are implemented for topological quantum magnets.

%% file: 1d-case.tex
\begin{figure}[t!]
    \centering
    \par\vspace{0.5em}
    \subfigure{
        \begin{overpic}[width=0.5\linewidth]{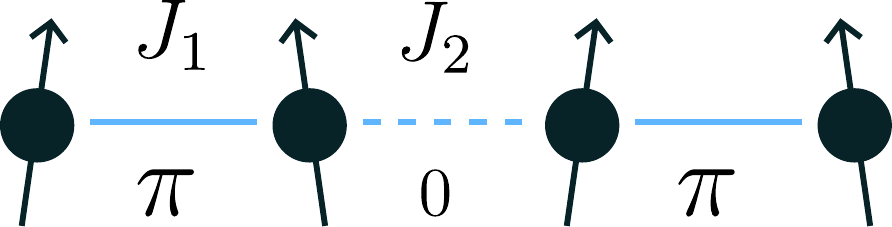}
            \put(-52,25){\small (a)}   
        \end{overpic}
    }
    \par\vspace{1em}
    \subfigure{
        \begin{overpic}[width=1.0\linewidth]{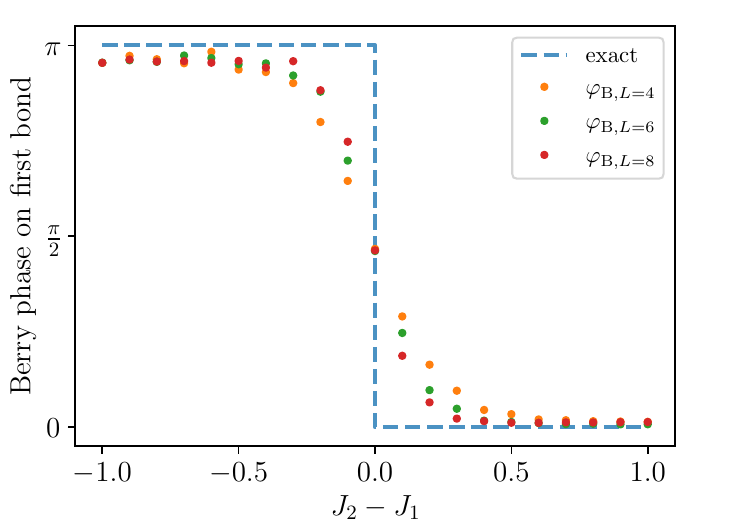}
            \put(-3,70){\small (b)}   
        \end{overpic}
    }
          
    \caption{(a) Schematic of a dimerized Heisenberg model featuring an SPT phase, with alternating bond strengths $J_1$ and $J_2$, and their respective Berry phase for $J_1 > J_2$. (b) We show the Berry phase on the first bond as a function of the interaction strength difference $J_2 - J_1$, with a comparison to the exact solution (in blue). With increasing system size $L$, convergence towards the exact solution improves. A phase transition can be seen where the Berry phase $\varphi_B$ changes from $\pi \rightarrow 0$, corresponding to topologically non-trivial and trivial states.}
    \label{fig:dimerized-results}
\end{figure}

The one-dimensional bond alternating spin-1/2 chain has been shown to host a rich phase diagram, including ferromagnetism, anti-ferromagnetism, Luttinger liquids, among others \cite{Luo2021}.
This system has been shown to reach a valence-bond ground state~\cite{Sachdev2008} with protected edge states~\cite{Watanabe2017}. Furthermore, dimerized systems can be realized in engineered systems, 
including spins on surfaces~\cite{wang2024realizing} and nano-graphene systems \cite{Zhao2024}. 
These systems can be described by the dimerized Heisenberg spin-1/2 model of the form

\begin{equation}
        H = \sum_{i} J_i \boldsymbol{S}_i \cdot \boldsymbol{S}_{i+1} = \sum_i H_i,
\end{equation}
where $J_{i} = \{J_1 \, \text{for odd} \, i,J_2 \, \text{for even} \, i \}$  is the alternating bond strength, and $\boldsymbol{S}_i$ is the vector of spin operators on lattice site $i$, whose entries are the Pauli matrices $\boldsymbol{S} = \frac{\hbar}{2} \left( \sigma_x, \sigma_y, \sigma_z \right)$. 
A bond can be gauged with a phase of the form~\cite{Hatsugai2006, Hatsugai2007}
\begin{equation}
    \label{eq:twist-def}
    H_i\rightarrow \frac{1}{2} S^+_iS^-_{i+1} e^{-i \varphi} +  \frac{1}{2} S^-_iS^+_{i+1} e^{i \varphi} + S_i^z S_{i+1}^z,
\end{equation}
where $S^\pm = \frac{\hbar}{2} \left(\sigma_x \pm \sigma_y\right)$ are the spin ladder operators. We make the phase time-dependent $\varphi(t)$ and evolve the resulting time-dependent Hamiltonian in a loop according to Eq. \ref{eq:half-time} to obtain the Berry phase $\varphi_B$. Stronger bonds yield $\varphi_B=\pi$ while weak bonds result in $\varphi_B=0$ (Fig.~\ref{fig:dimerized-results}a). The topological difference becomes evident when removing a strong bond, the resulting open boundary chain exhibits edge modes \cite{lado2019}. If a weak bond is removed, no edge modes appear. In the case of $J_1=J_2$ the system becomes degenerate during the evolution and no Berry phase can be defined \cite{Xiao2010}.

We build the quantum circuit by first decomposing the Hamiltonian into time-dependent and time-independent parts $H(t) = H_0 + H_1(t)$. If the twist operation is on an even (odd) bond, then the time independent part $H_0$ will contain all odd (even) bonds. The time-dependent part contains only one non-commuting term at the twisted bond of the form $H_1'(t) \propto \sigma^x \otimes \sigma^y - \sigma^y \otimes \sigma^x$, so we split the time-dependent part $H_1(t) = H_1'(t)+H_1''(t)$, with $H_1''(t)$ having the rest of the commuting terms. The decomposed Hamiltonian $H(t)=H_0+H_1'(t)+H_1''(t)$ is then used to approximate the time evolution as a second-order Trotter-Suzuki decomposition (cf. Eq. ~\ref{eq:trotter-suzuki}). If the twist is specified at the even bond $j$, then the corresponding unitary operators are

\begin{figure*}[t!]
    \centering
    \raisebox{-0.2cm}{\begin{quantikz}[column sep=0.6cm, row sep=0.4cm]
        & \gate[4]{U_2(t + \Delta t)} & \\
        & \qw \vphantom{\frac{\Delta t}{2}} & \\
        & \qw \vphantom{\frac{\Delta t}{2}} & \\
        &  \qw \vphantom{\frac{\Delta t}{2}} & \\
    \end{quantikz}}
    =
    \begin{quantikz}[column sep=0.6cm, row sep=0.4cm]
    & \gate[2]{J_1} 
      \gategroup[4,steps=1,style={dashed,rounded corners,inner sep=2pt}]{\small $H_0$}
    & 
      \gategroup[4,steps=1,style={dashed,rounded corners,inner sep=2pt}]{\small $H_1'(t)$}
    & 
      \gategroup[4,steps=1,style={dashed,rounded corners,inner sep=2pt}]{\small $H_1''(t)$}
    & 
      \gategroup[4,steps=1,style={dashed,rounded corners,inner sep=2pt}]{\small $H_1'(t)$}
    & \gate[2]{J_1} 
      \gategroup[4,steps=1,style={dashed,rounded corners,inner sep=2pt}]{\small $H_0$}
    & \\
    & \qw
    & \gate[2]{J_2 \sin \varphi(t + \frac{\Delta t}{2})}
    & \gate[2]{J_2 \cos \varphi(t + \frac{\Delta t}{2})}
    & \gate[2]{J_2 \sin \varphi(t + \frac{\Delta t}{2})}
    & \qw
    & \\
    & \gate[2]{J_1}
    & \qw
    & \qw
    & \qw
    & \gate[2]{J_1}
    & \\
    & \qw
    & \qw
    & \qw
    & \qw
    & \qw
    &
    \end{quantikz}
    \caption{Quantum circuit diagram of the general form of the implemented second-order Trotter-Suzuki unitary time-step $U_2(t+\Delta t)$. The full unitary evolution will be given as a product of these small time steps.}
    \label{fig:unitary-circuit}
\end{figure*}

\begin{equation}
    \label{eq:unitary-h0}
    e^{-\frac{i}{\hbar} \frac{\Delta t}{2} H_0} = \prod_{i \, \in \,\text{odd}} 
    e^ {-\frac{i}{\hbar}  \frac{\Delta t}{2} J_1 \, \boldsymbol{S}_i \, \cdot \, \boldsymbol{S}_{i+1} },
\end{equation}

\begin{align}
    \label{eq:unitary-h1-1}
    &e^{-\frac{i}{\hbar} \frac{\Delta t}{2} H_1'(t + \frac{\Delta t}{2})} = 
    \\
     &e^{\frac{i}{\hbar}  \frac{\Delta t}{2} J_2 \, 
      \sin \varphi\left(t + \frac{\Delta t}{2} \right)
      \left( 
     S^x_j S^y_{j+1} - S^y_j S^x_{j+1}
     \right)
     }, 
    \notag 
    \\
    \notag
    \\
    \label{eq:unitary-h1-2}
    &e^{-\frac{i}{\hbar} \frac{\Delta t}{2} H_1''(t + \frac{\Delta t}{2})} = 
    \\
    &e^{-\frac{i}{\hbar} \frac{\Delta t}{2} J_2 \, 
        \cos \varphi\left(t + \frac{\Delta t}{2} \right)
        \left( 
        S^x_j S^x_{j+1} + S^y_j S^y_{j+1}
        \right)
     } 
    \notag 
    \\
    &\times e^{-\frac{i}{\hbar} \frac{\Delta t}{2} J_2 \, 
    S^z_j S^z_{j+1}
    }
    \prod_{i \, \in \, \text{even} \, \ne \, j} 
    e^{-\frac{i}{\hbar}  \frac{\Delta t}{2} J_2 \, \boldsymbol{S}_i \, \cdot \, \boldsymbol{S}_{i+1}} .
    \notag 
\end{align}

We implement the previous decomposition in Qiskit using only two qubit gates (details in App. \ref{appendix:qiskit-gates}) \cite{zenodoQC}. We outline the constituents of the quantum circuit as a diagram in Fig. ~\ref{fig:unitary-circuit}, for a single trotterized unitary time step. Furthermore, Qiskit includes methods to create a controlled gate given an arbitrary unitary operator \cite{qiskit_controlledgate}, which we exploit to implement the Hadamard test of the trotteirzed unitary evolution as described in the methods. We implement this circuit using Qiskit's AerSimulator \cite{qiskit} with 100~000 shots. Ground state initialization was achieved with Qiskit's built-in state preparation method. A parameter sweep was performed for different dimer strengths $J_2$, setting $J_1 = 1$, $T = 20$, $N = 200$, and $\hbar = 1$. System sizes $L = \{4, 6, 8\}$ were tested as well. The results are plotted in Fig.~\ref{fig:dimerized-results}b. 

A phase transition can be seen at a bond when changing the relative strengths of the alternating pattern. The topological order parameter changes between $0$ and $\pi$, corresponding to trivial and non-trivial phases. Errors increase when the link strengths approach each other and finite size effects become observable, yielding stronger deviations around $J_2-J_1=0$ from the exact value. Although the circuit shows promise, the number of gates and circuit depth ranged in the order of $\mathcal{O}(10^5)$. Due to the large circuit, error mitigation schemes are not implemented. It is worth noting that due to the required depth, this circuit is not runnable on currently available quantum computers, requiring either substantially higher qubit fidelities, topological qubits or full error correction~\cite{Preskill2018}.

%% file: 2d-case.tex
\begin{figure}[t!]
    \centering
    \includegraphics[width=0.8\linewidth]{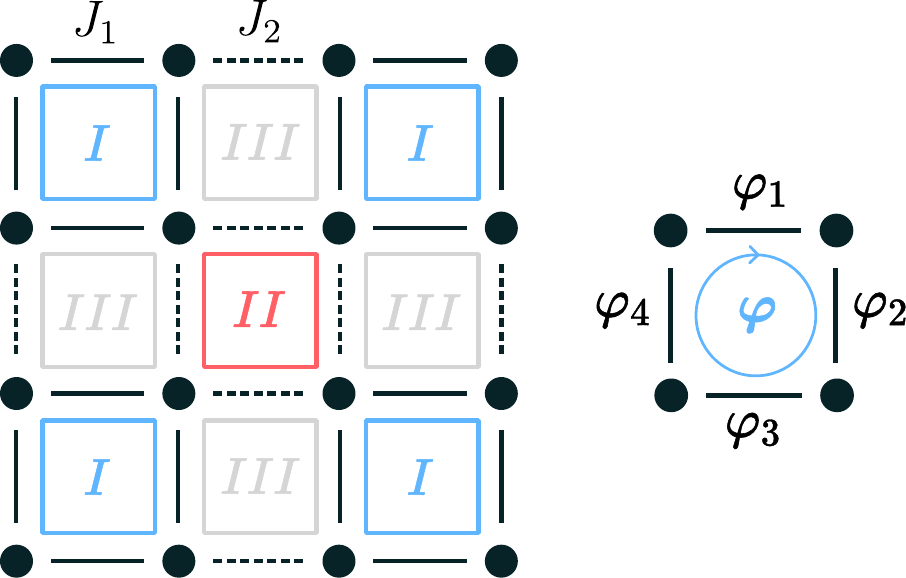}
    \caption{Illustration of a two-dimensional tetramerized Heisenberg spin model and gauged twists, on an exemplary $4\times 4$ system. The alternating interaction strengths $J_1$ and $J_2$ are indicated by solid and dashed lines, respectively. This system exhibits a plaquette structure. Plaquettes of type I contain only nearest neighbor interactions of strength $J_1$,  plaquettes of type II consist of strengths $J_2$, and plaquettes of type III are mixed. In order to calculate the Berry phase, we gauge a plaquette with four twists $\varphi_j$ on each bond.}
    \label{fig:tetra-model}
\end{figure}

We now study the two-dimensional SPT phase, a frustrated tetramerized square lattice spin-1/2 Heisenberg model. This model exhibits a rich phase diagram, including long-range order due to spontaneous symmetry breaking as well as higher-order topology ~\cite{GonzlezCuadra2022,PhysRevResearch.7.013194,wang2024realizing}. 
This HOSPT phase has been experimentally realized with atomically engineered quantum spin lattices,
including their topological corner modes \cite{wang2024realizing}. 
The Hamiltonian takes the form

\begin{equation}
        H = \sum_{\langle i,j \rangle} J_{ij} \boldsymbol{S}_i \cdot \boldsymbol{S}_{j},
\end{equation}
where the summation is over nearest neighbors and the interaction strengths $J_{ij}$  have the valence-bond pattern with strengths $J_1$ for plaquettes of type~I and $J_2$ for plaquettes of type~II as shown in Fig.~\ref{fig:tetra-model}. Plaquettes of type~III have mixed strengths, which are completely defined by the pattern of types~I and II.

This model hosts a higher-order symmetry protected topological phase. The topological invariant for these types of phases are the $\mathbb{Z}_Q$ invariants~\cite{Araki2020}. In the case of  HOSPT phases with anti-unitary symmetry, the Berry phase invariant is still quantized to $\left\{0,\pi \right\}$. We can proceed analogously as in the previous section by adding a twist on each bond of the plaquette (cf.~Eq.~\ref{eq:twist-def}) \cite{Araki2020}. To find valid parameter paths, we turn to the decoupled case of $J_2 = 0$, where the Hamiltonian becomes a sum over decoupled plaquettes. In this decoupled case the four lowest energy levels of the spectrum of a single plaquette (procedure outlined in App. ~\ref{appendix:degen}) are of the form

\begin{equation}
    \label{eq:degen-energy}
    E_{\text{min}}=-2J_1\pm2J_1 \left[5 \pm 4 \left| \cos \frac{|\boldsymbol{\varphi}|}{2} \right| \right]^{1/2},
\end{equation}
where the twists are grouped into a vector $\boldsymbol{\varphi} =\left(\varphi_1, \varphi_2, \varphi_3, \varphi_4 \right)$, such that the parameter vector distance is $|\boldsymbol{\varphi}|= \sum_j \varphi_j$. One can see that the ground state becomes degenerate whenever $|\boldsymbol{\varphi}|= \pi$. Thus, valid parameter paths are loops in a 4-torus parameter space such that the distance is always $|\boldsymbol{\varphi}(t)| < \pi$. A simple way to achieve this is to evolve parameters in opposite pairs $\varphi_i(t) = -\varphi_j(t)$, such that the vector distance is always canceled $|\boldsymbol{\varphi}(t)| = 0$, ensuring that if the system starts in the ground state, it will always stay there during evolution.

Defining plaquettes of type~I as those which have stronger interaction strength $J_1$ as compared to plaquettes of type~II with weaker interaction $J_2$, twisting plaquettes of type~I yields $\varphi_B=\pi$ and of type~II yields $\varphi_B=0$ \cite{GonzlezCuadra2022, Araki2020}. The topological inequivalence follows analogously to the one-dimensional case. By "cutting" through plaquettes and considering the resulting open boundary system, splitting a type~I plaquette will give corner modes while a type~II will not \cite{Araki2020, GonzlezCuadra2022}.

We build the quantum circuit by dividing the Hamiltonian into the time-dependent part (which contains the twists) and time-independent part, such that $H(t) = H_0+H_1(t)$. If the twist is on a plaquette of type I (type II), the time-independent part $H_0$ will then consist of all $J_2$ ($J_1)$ interactions, while the time-dependent part $H_1(t)$ contains all $J_1$ ($J_2$) interactions. These resulting parts are then split into vertical and horizontal interactions such that $H_0 = H_{0,V} + H_{0, H}$, $H_1(t) = H_{1,V}(t) + H_{1, H}(t)$. Finally, both time-dependent parts are split by their non-commuting operators as in the previous section such that $H_{1,\alpha}(t) = H'_{1,\alpha}(t) + H''_{1,\alpha}(t)$, with $\alpha = \left\{V,H\right\}$. This partition of the Hamiltonian is used to approximate the time evolution as a second-order Trotter-Suzuki decomposition (cf. Eq. \ref{eq:trotter-suzuki}). The unitary operators are identical to their 1D counterparts (Eq. ~\ref{eq:unitary-h0} - \ref{eq:unitary-h1-2}).

\begin{figure}[t]
    \centering
    \includegraphics[width=0.9\linewidth]{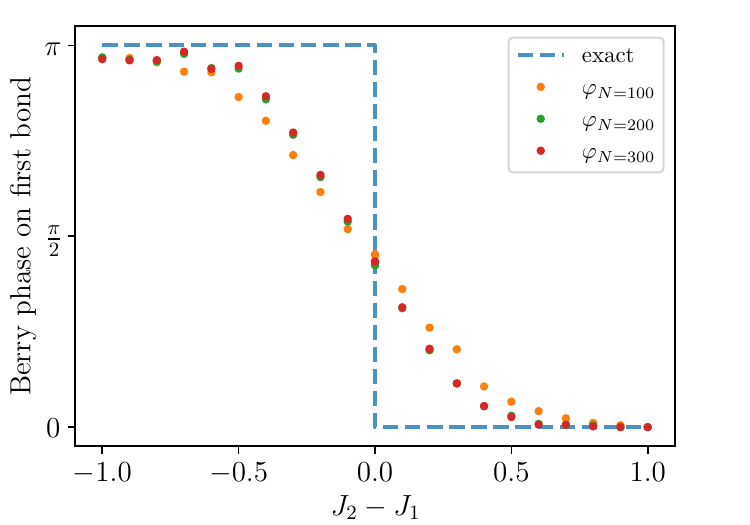}
    \caption{
    Berry phase calculation on a plaquette as a function of the interaction strength difference $J_2 - J_1$, with a comparison to the exact solution (in blue). With increasing time steps $N$, convergence towards the exact solution improves. A phase transition can be seen where the Berry phase $\varphi_B$ changes from $\pi \rightarrow 0$, corresponding to higher-order topologically non-trivial and trivial states.
    }
    \label{fig:tetra-results}
\end{figure}

\begin{figure}[t]
    \centering
     \begin{overpic}[width=1\linewidth]{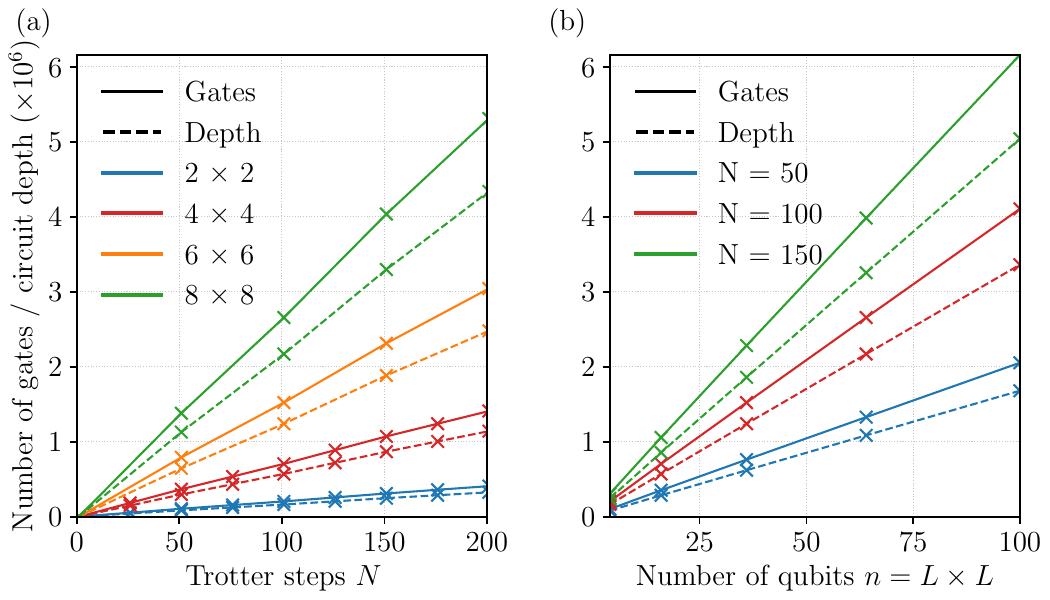}
    \end{overpic}
    \caption{
    (a) Gates and circuit depth counts for the two-dimensional quantum circuit on $L \times L$ spin systems as a function of trotter steps $N$. Linear dependence on $N$ is due to each time step being fundamentally the same circuit $U_2(t+\Delta t)$. (b) Gates and depth counts as a function of number of qubits $n = L\times L$ at $N=100$. Linear dependence on $n$ is fundamentally due to the model's nearest-neighbor interactions and decomposition into two qubit rotations as described in the text.}
    \label{fig:tetra-gates-depths}
\end{figure}

We implement this circuit using the same procedure as in the previous section. We demonstrate this circuit for a $4 \times 4$ spin system \cite{zenodoQC}, using Qiskit's AerSimulator ~\cite{qiskit} with 100~000 shots. A parameter sweep was performed for different dimer strengths $J_2$, setting $J_1 = 1$, $T = 20$, and $N =\left\{100, 200, 300 \right\}$, and results are shown in Fig.~\ref{fig:tetra-results}. 
For the case of $N=300$ steps, gate counts are in the order of $\mathcal{O}(10^6)$, which cannot be directly implemented in currently available quantum computers. The number of gates and circuit depth have a linear dependence on the number of trotter steps $N$, as well as for number of qubits $n=L \times L$ as shown in Fig. ~\ref{fig:tetra-gates-depths}. This is due to the model having only nearest-neighbor interactions. Without double counting, the model requires implementing $2n$ interactions per time step, which means implementing $6n$ two-qubit rotations. Both conversion into a controlled rotation with an auxiliary qubit, as well as using only a universal set of gates, gives overhead which is linear with the number of gates \cite{Nielsen2012} such that the final number of gates per time step is $G(n) \sim \mathcal{O}(n)$.

An interesting finding is that not all parameter paths yield the expected Berry phase. When only evolving two twists in a type I plaquette (leaving the other two stationary), we observed that twisting commuting links (i.e their spin bonds are not directly connected / interacting), the Berry phase would yield $0$ rather than $\pi$. We summarize these findings in Table~\ref{tab:weird-tetra}. This result highlights an important feature, a higher-dimensional parameter geometry does not necessarily give the same topological charge when traversing it along different closed paths. Gauging a symmetry and twisting the parameters can become non-trivial and one needs to carefully evolve through the introduced parameter space. 

\begin{table}[t]
\begin{center}
\begin{tabular}{||c c c||} 
 \hline
 Twisted pair & Berry phase & Commute \\ [0.5ex] 
 \hline\hline
 $\varphi_1$, $\varphi_2$ & $\pi$ & No \\ 
 \hline
 $\varphi_1$, $\varphi_4$ & $\pi$ & No  \\
 \hline
 $\varphi_2$, $\varphi_3$ & $\pi$ & No  \\
 \hline
 $\varphi_3$, $\varphi_4$ & $\pi$ & No  \\
 \hline
 $\varphi_2$, $\varphi_4$ & $0$& Yes \\
 \hline
 $\varphi_1$, $\varphi_3$ & $0$ & Yes  \\ [1ex] 
 \hline
\end{tabular}
\caption{Berry phase for a type~I plaquette, only twisting two links. Twists on interactions that commute in the Hamiltonian do not yield the expected phase.}
\label{tab:weird-tetra}
\end{center}
\end{table}